\documentclass[article, 10pt]{article}
\usepackage{amsmath}
\usepackage{amsthm}
\usepackage[dvips]{graphicx}
\usepackage{wrapfig}
\usepackage{pxfonts}
\usepackage{bm}
\usepackage[font={small,sl}]{caption}

\newtheoremstyle{physics}{}{}{\itshape}{}{\bfseries}{}{\newline}{}

\theoremstyle{physics}

\renewcommand\d{\mathrm{d}}

\begin{document}

\title{Black holes and singularities \\ in causal set gravity}
\author{Yu Asato\thanks{email: yasato@ucdavis.edu} \\ {\it Department of Physics} \\ {\it University of California} \\ {\it Davis, CA 95616} \\ {\it USA}}
\date{May, 2019}
\maketitle

\begin{abstract}
  A precise definition of a black hole has long been absent in causal set theory.
  I first show that the local finiteness of the theory cannot be interpreted as a complete discretization condition, and the theory still admits continua, which I call singular anti-chains.
  After providing a couple of suggestive calculations from general relativity to support the argument that a singular anti-chain behaves like a causal singularity in a space-time, I propose a definition of a causal set black hole based on a singular anti-chain.
  The paper concludes with brief discussions of questions that causal set black holes and singularities may give rise to.\\[10pt]
\end{abstract}

\section{Introduction}

A causal set (or causet) is a partially ordered set with an extra condition called local finiteness.
That is, a pair consisting of a set and a binary operator, $(C, \preceq)$, is a causal set if it satisfies (i) $x\preceq y\preceq z \Rightarrow x\preceq z$ (transitivity), (ii) $x\preceq y$ and $ y\preceq x \Rightarrow x=y$ (acyclicity), and (iii) $|A(x,y)|<\infty$ (local finiteness), where $x, y, z$ are arbitrary elements of $C$, $A(x,y):=\{z\in C~|~x\preceq z \preceq y \}$ is an Alexandroff interval, and $|\cdot |$ denotes the cardinality of the set.
Causal set theory is one of the approaches to quantum gravity, in which a causet plays a prominent role as a space-time, and the partial order is considered to be the causal order of the spacetime \cite{BLMS, Sorkin}.

Causal set theory has many intriguing aspects as a theory of discrete space-time:
We can deduce time-like and space-like distances from it \cite{ITR, RW}; it maintains the Lorentz symmetry \cite{BHS}; quantum fields can be defined on causets \cite{Johnston, Sverdlov}; the Benincasa-Dowker action acts as a causal set action \cite{BD}, etc.
However, one important concept that is still unclear in the theory is how to describe black holes.
Although there exist some important works on black holes in causal set theory \cite{DS, RZ, HR}, they have relied on the sprinkling process, in which we utilize a pre-existing black hole space-time as a background to produce a causet, and a precise definition of a causal set black hole is missing.

In the following, I first point out that the theory is not completely discretized; it still admits the existence of continua.
I then discuss a possible interpretation of this uncountable set in the theory as a causal singularity, and provide suggestive calculations from general relativity to support the argument.
Based on the causal set singularity, I propose a definition of a causal set black hole.

\section{Continuum in causal set theory} \label{sec:cont}

\begin{figure}[h] \centering
  \begin{picture}(300,50)    
    \linethickness{1.5pt}
    \put(75,5){\line(0,1){40}}
    \put(175,25){\line(1,0){100}}    
      
    \linethickness{0.2pt}
    \put(150,0){\line(0,1){50}}
  \end{picture}
  \caption{Uncountable chain and uncountable anti-chain (singular anti-chain). Left: All elements are related, and cause violation of local finiteness. Right: No elements are related, and there is no violation of local finiteness.}
  \label{fig:cont}
\end{figure}
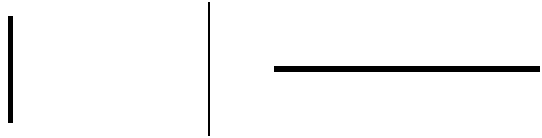

We start by reflecting on the local finiteness condition.
We usually consider the condition to be the discretization condition of the theory \cite{Sorkin}.
Since local finiteness states that local regions defined by Alexandroff intervals, or causal diamonds, must only have a finite number of elements, we think any causet should have, at most, a countable number of elements.
More precisely, we tend to believe the following statement:
\begin{equation} \label{eq1}
  \forall ~x, y \in C,~|A(x,y)|<\infty \quad \overset{?}{\Rightarrow} \quad |C|\leqslant \aleph _0~,
\end{equation}
and we interpret the local finiteness condition as the discretization condition since any causet is a countable set due to the local finiteness. 
However, the statement (\ref{eq1}) is not true.
The local finiteness condition is not restrictive enough to exclude all continua from the theory.
While it certainly rules out the possibility of having ``temporal'' continua, or uncountable chains, ``spatial'' continua, or uncountable anti-chains, are still contained in the theory (see Figure \ref{fig:cont}).
Here a chain is a totally ordered subset of a causet, and an anti-chain is a totally unordered subset of a causet.
We call an anti-chain \emph{singular} if it is uncountable, and a causet \emph{singular} if it contains singular anti-chains.

Now the question is whether there exists any physical interpretation of singular anti-chains, or their existence is a mere loophole of the theory.
In the following, I seek to provide a possible interpretation of a singular anti-chain as a causal singularity in a space-time. 
In order to explain this, I define a \emph{classical causality} between an element, $x\in C$, and a singular anti-chain, $S$:
\begin{equation}
  x\preceq S\quad \overset{\mbox{\scriptsize def.}}{\Longleftrightarrow }\quad \forall ~s\in S,~x\preceq s~.
\end{equation}
In a sense, we regard a singular anti-chain as a single spacetime element that is ``fatter'' than usual elements.
Thus, it is more natural to consider a collection of causalities between a non-singular element and elements of a singular anti-chain as a single causality, which is a classical causality.
We can certainly think about an additional, finite number of causalities involving a singular anti-chain, but it is beyond the scope of the paper, and this possibility will be discussed elsewhere.

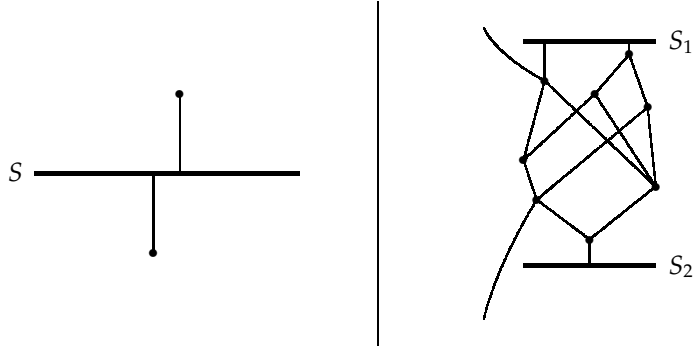
\begin{figure}[h] \centering
  \begin{picture}(300,130)    
    \linethickness{1.5pt}
    \put(20,65){\line(1,0){100}}
    \put(10,62){$S$}
    \linethickness{0.4pt}

    \put(65,35){\circle*{3}}
    \put(75,95){\circle*{3}}

    \qbezier(65,35)(65,35)(65,65)

    \qbezier(75,95)(75,95)(75,65)
    
    \linethickness{0.2pt}
    \put(150,0){\line(0,1){130}}

    \linethickness{1.5pt}
    \put(205,115){\line(1,0){50}}
    \put(260,112){$S_1$}
    \put(205,30){\line(1,0){50}}
    \put(260,27){$S_2$}
    \linethickness{0.4pt}

    \put(230,40){\circle*{3}}
    \put(210,55){\circle*{3}}
    \put(255,60){\circle*{3}}
    \put(205,70){\circle*{3}}
    \put(252,90){\circle*{3}}
    \put(213,100){\circle*{3}}
    \put(232,95){\circle*{3}}
    \put(245,110){\circle*{3}}

    \qbezier(230,40)(230,40)(210,55)
    \qbezier(230,40)(230,40)(255,60)
    \qbezier(210,55)(210,55)(205,70)
    \qbezier(210,55)(210,55)(252,90)
    \qbezier(255,60)(255,60)(252,90)
    \qbezier(205,70)(205,70)(213,100)
    \qbezier(205,70)(205,70)(232,95)
    \qbezier(232,95)(232,95)(245,110)
    \qbezier(252,90)(252,90)(245,110)
    \qbezier(255,60)(255,60)(232,95)
    \qbezier(255,60)(255,60)(213,100)

    \qbezier(213,100)(195,110)(190,120)
    \qbezier(210,55)(195,30)(190,10)

    \qbezier(230,40)(230,40)(230,30)

    \qbezier(245,110)(245,110)(245,115)
    \qbezier(213,100)(213,100)(213,115)
  \end{picture}
  \caption{Examples of singular causets. $S$ denotes a singular anti-chain. Solid lines connecting a singular anti-chain and an element represent classical causalities unless an element in the singular anti-chain is specified. Left: A singular causet violating local finiteness. Right: An allowed singular causet having two singular anti-chains.}
  \label{fig:sac}
\end{figure}

What is interesting about singular anti-chains and classical causalities is that a singular anti-chain cannot have the classical future and the classical past at the same time.
That is, a singular anti-chain behaves like a causal singularity in our classical spacetime.
If we consider a singular anti-chain that has an element to its future and another element to its past both of which are related to the singular anti-chain by classical causalities, then we can clearly see the violation of the local finiteness by selecting those two non-singular elements and forming a causal diamond (see Figure \ref{fig:sac}).

In this view, non-singular regions of a classical spacetime are still discretized in a standard causal set manner, but singular causal boundaries refuse to be discretized, and continue to be continuous.

\section{Suggestive signs from general relativity} \label{sec:grcalc}

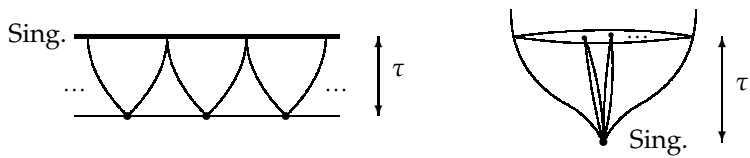
\begin{figure}[h] \centering
  \begin{picture}(300, 70)
    \linethickness{1.5pt}
    \put(25,50){\line(1,0){100}}
    \linethickness{0.4pt}
    \put(0,48){Sing.}
    \put(25,20){\line(1,0){100}}
    \put(75,20){\circle*{3}}
    \put(105,20){\circle*{3}}
    \put(45,20){\circle*{3}}
    \qbezier(75,20)(90,35)(90,50)
    \qbezier(75,20)(60,35)(60,50)
    \qbezier(45,20)(60,35)(60,50)
    \qbezier(45,20)(30,35)(30,50)
    \qbezier(105,20)(90,35)(90,50)
    \qbezier(105,20)(120,35)(120,50)
    \multiput(22,30)(3,0){3}{\circle*{1}}
    \multiput(121,30)(3,0){3}{\circle*{1}}

    \put(140,20){\vector(0,1){30}}
    \put(140,50){\vector(0,-1){30}}
    \put(145,35){$\tau$}

    \qbezier(225,10)(220,20)(210,25)
    \qbezier(225,10)(230,20)(240,25)
    \qbezier(210,25)(190,37)(190,60)
    \qbezier(240,25)(260,37)(260,60)

    \qbezier(191,50)(230,55)(259,50)
    \qbezier(191,50)(220,45)(259,50)

    \put(225,10){\circle*{3}} \put(235,8){Sing.}

    \put(228,50.5){\circle*{2}}
    \put(218,49.5){\circle*{2}}
    \multiput(235,50)(3,0){3}{\circle*{1}}

    \qbezier(225,10)(225,30)(228,50.5)
    \qbezier(225,10)(230,30)(228,50.5)
    \qbezier(225,10)(220,29)(218,49.5)
    \qbezier(225,10)(225,29)(218,49.5)

    \put(270,10){\vector(0,1){40}}
    \put(270,50){\vector(0,-1){40}}
    \put(275,30){$\tau$}
  \end{picture}
  \caption{Configurations of embedded anti-chains. The left is for the Schwarzschild black hole, and the right is for the FLRW universe.}
  \label{fig:signs}
\end{figure}

In this section, we investigate whether there are any signs of singular anti-chains in general relativity.
To explore this, we calculate the minimum density of elements of an anti-chain embedded near a singularity so that its future or past covers the entire singularity as shown in Figure \ref{fig:signs}.
More precisely, we place a 3-hypersurface near the singularity, and embed an anti-chain uniformly into the hypersurface so that the future or the past of the anti-chain covers the entire singularity.
We calculate 3-volume that a single embedded element occupies on the hypersurface, which can be interpreted as the ``discreteness scale'' of the anti-chain.
The inverse of this 3-volume is the density we are interested in, and it is equivalent to the number of embedded anti-chain elements per unit 3-volume.
We then send the proper time between the hypersurface and the singularity to zero, and observe the behavior of the density.
If singular anti-chains correspond to singularities, the density should not converge because a converging density implies a discrete anti-chain along the singularity.

\subsection{Schwarzschild black hole}

We first consider the Schwarzschild solution in standard Schwarzschild coordinates:
\begin{equation} \label{eq3}
  g_{ab} = -\left( 1-\frac{r_{\mathrm{S}}}{r} \right) \d t^2+\left( 1-\frac{r_{\mathrm{S}}}{r} \right)^{-1} \d r^2+r^2 \d \Omega^2 ~,
\end{equation}
where $r_{\mathrm{S}} = 2GM$ is the Schwarzschild radius, and $\d \Omega ^2 = \d \theta ^2 + \sin^2\theta \d \phi ^2$.
A null geodesic in the Schwarzschild space-time must entirely stay on a single plane defined by the position and the velocity vectors because of the parity reflection symmetry.
We can orient the coordinate system so that this plane becomes defined by $\theta = \pi/2$, and the conserved quantities of the geodesic can be written as
\begin{equation} \label{eq4}
  E=\left( 1-\frac{r_{\mathrm{S}}}{r} \right) \frac{\d t}{\d \lambda}, \quad L=r^2\frac{\d \phi}{\d \lambda}~,
\end{equation}
where $\lambda $ is an affine parameter.
Using eqs.~(\ref{eq3}) and (\ref{eq4}), we then obtain
\begin{equation}
  \left(\frac{\d r}{\d \lambda}\right)^2 + V =0~, \quad \mbox{where} \quad V=\frac{L^2}{r^2}-\frac{r_{\mathrm{S}}L^2}{r^3}-E^2~,
\end{equation}
for the null geodesic \cite{Wald}.

\begin{figure}[h]\centering
	\includegraphics [width=10cm,clip]{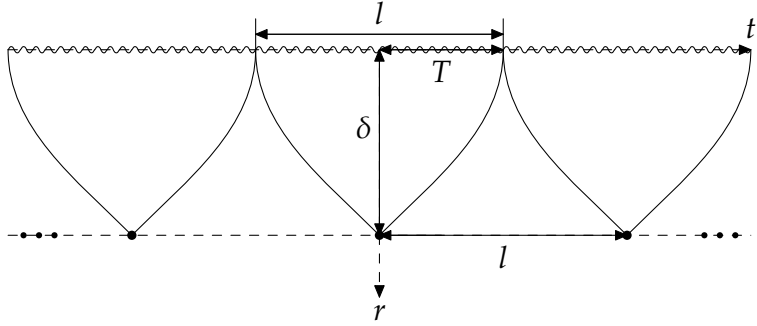}
	\caption{A $t$-$r$ diagram showing the configuration of embedded anti-chain elements near the Schwarzschild singularity. The separation between two embedded elements is the same as the distance on the singularity covered by a future of a single element.}
	\label{fig:schw}
\end{figure}

Now we consider an anti-chain uniformly embedded in the hypersurface constant coordinate time $\delta$ away from the singularity as shown in Figure \ref{fig:schw}.
Since we are interested in the case where $\delta \ll r_{\mathrm{S}}$, the $r$-coordinate works as a time coordinate now inside the event horizon ($r < r_{\mathrm{S}}$).
The union of the future light cones of anti-chain elements covers the singularity, but we want to maximize the $t$-coordinate separation between two embedded elements in order to find the lower bound of the density of the anti-chain elements.
When they are maximally separated, the separation is equal to the $t$-coordinate distance on the singularity that is covered by a future of a single embedded element.
Thus, in the following, we calculate the maximum $t$-coordinate distance light travels from the hypersurface to the singularity.
   
Let's first think about the case where $L=0$. We have $V=-E^2$, and
\begin{equation}
  \frac{\d r}{\d \lambda} = \pm E \quad \Leftrightarrow \quad   \frac{\d r}{\d t} = \pm \left(1-\frac{r_{\mathrm{S}}}{r}\right)~.
\end{equation}
We obtain
\begin{eqnarray}
  \pm T & = & \pm \int ^T _0 \d t ~=~ \int ^0 _\delta \left(1-\frac{r_{\mathrm{S}}}{r}\right)^{-1} \d r \notag \\
        & = & \int ^\delta _0 \frac{r}{r_{\mathrm{S}}}\left(1-\frac{r}{r_{\mathrm{S}}}\right)^{-1} \d r ~=~ -\delta -r_{\mathrm{S}}\ln \left(1-\frac{\delta}{r_{\mathrm{S}}}\right) \notag \\
        & \sim & \frac{\delta ^2}{2r_{\mathrm{S}}} ~.
\end{eqnarray}
Here we used $\delta \ll r_{\mathrm{S}}$.
Thus, when $L=0$, the $t$-coordinate distance covered on the singularity becomes $l_0\equiv 2T\sim \delta ^2 / r_{\mathrm{S}}$ .

Let's now consider the case where $L\neq 0$.  The leading term of $V$ becomes $-\frac{r_{\mathrm{S}}L^2}{r^3}$, and we have
\begin{equation}
  \frac{\d r}{\d \lambda} = \pm \frac{\sqrt{r_{\mathrm{S}}}L}{r^{3/2}} \quad \Leftrightarrow \quad \frac{\d r}{\d t} = \pm \sqrt{r_{\mathrm{S}}} \cdot \frac{L}{E} \cdot r^{-3/2}\left(1-\frac{r_{\mathrm{S}}}{r}\right)~.
\end{equation}
We, in this case, obtain
\begin{eqnarray}
	  \pm \sqrt{r_{\mathrm{S}}} \cdot \frac{L}{E} \cdot T & = & \pm \sqrt{r_{\mathrm{S}}} \cdot \frac{L}{E} \int ^T _0 \d t 
	  = \int ^0 _\delta r^{3/2}\left( 1-\frac{r_{\mathrm{S}}}{r} \right)^{-1} \d r \notag \\
	& = & \frac{1}{r_{\mathrm{S}}} \int ^\delta _0 r^{5/2}\left( 1-\frac{r}{r_{\mathrm{S}}} \right)^{-1} \d r 
	\sim \frac{1}{r_{\mathrm{S}}} \int ^\delta _0 r^{5/2}\left( 1+\frac{r}{r_{\mathrm{S}}} \right) \d r \notag \\
	& = & \frac{2\delta ^{7/2}}{7r_{\mathrm{S}}} + \frac{2\delta ^{9/2}}{9r_{\mathrm{S}} ^2}
	\sim \frac{2\delta ^{7/2}}{7r_{\mathrm{S}}}~,
\end{eqnarray}
\begin{equation}
  \therefore \quad \pm T \sim \frac{2E\delta^{7/2}}{7r_{\mathrm{S}}^{3/2}L}~.
\end{equation}
Thus, when $L\neq 0$, the $t$-coordinate distance covered on the singularity becomes $l_L \equiv 2T \sim \frac{4E\delta^{7/2}}{7r_{\mathrm{S}}^{3/2}L}$.

We notice that $l_0$ is larger than $l_L$:
\begin{equation}
  l_L = \frac{4}{7}\cdot\frac{E}{L}\cdot\frac{\delta^{7/2}}{r_{\mathrm{S}}^{3/2}} \ll \frac{4}{7}\cdot\frac{1}{\delta}\cdot\frac{\delta^{7/2}}{r_{\mathrm{S}}^{3/2}}=\frac{4}{7}\cdot\left(\frac{\delta}{r_{\mathrm{S}}}\right)^{1/2}\cdot l_0 \ll l_0~.
\end{equation}
Here we used the fact that we can always choose $\delta$ such that $E/L \ll 1/\delta$.
Hence, $l_0$ gives the maximum $t$-coordinate separation between two embedded anti-chain elements on the hypersurface.

We then convert the calculated coordinate distance and coordinate time into proper distance and proper time.
The proper distance is calculated along the straight line defined with constant $r$ and constant angles, and the proper time is calculated along the straight line defined with constant $t$ and constant angles:
\begin{gather}
  \mathcal{L} = \int ^\mathcal{L} _0 \d \mathcal{L} = \int ^l _0 \sqrt{\frac{r_\mathcal{S}}{\delta} - 1}~ \d l = l\sqrt{\frac{r_\mathcal{S}}{\delta} - 1} \sim l \sqrt{\frac{r_\mathcal{S}}{\delta}}~, \\
  \tau =\int ^\tau _0 \d \tau =\int ^\delta _0 \left( \frac{r_{\mathrm{S}}}{r} -1 \right)^{-1/2}\d r \sim \int ^\delta _0 \left[\left(\frac{r}{r_{\mathrm{S}}}\right)^{1/2} + \frac{1}{2} \left( \frac{r}{r_{\mathrm{S}}}\right)^{3/2} \right]\d r \sim \frac{2\delta ^{3/2}}{3r_{\mathrm{S}}^{1/2}}~.
\end{gather}
Hence, the one-dimensional density is given by
\begin{gather}
  \mathcal{L}_0 = l_0 \sqrt{\frac{r_\mathcal{S}}{\delta}} = \frac{\delta ^{3/2}}{r_{\mathrm{S}}^{1/2}}=\frac{3}{2}\tau~, \\
  \therefore \quad \rho _\mathrm{1-dim}\equiv \frac{1}{\mathcal{L}_0} =\frac{2}{3\tau}~.
\end{gather}

In order to obtain full three-dimensional expression of the density, we remember that multiple elements can be put at the same $t$ and $r$, but different angles, and we have to calculate the maximum angle separation, and the corresponding solid angle.

In coordinate time, we have
\begin{equation}
  \frac{\d \phi}{\d r} = \frac{\pm 1}{\sqrt{(E/L)^2r^4-r^2-r_{\mathrm{S}}\cdot r}}~,
\end{equation}
but since $\frac{E}{L}\ll \frac{1}{\delta}$, it approximately becomes
\begin{equation}
  \frac{\d \phi}{\d r} \sim \frac{\pm 1}{\sqrt{r_{\mathrm{S}}\cdot r}}~.
\end{equation}
Thus, as light travels to the singularity, it maximally sweeps the angle of
\begin{equation}
  \Delta \phi \sim \frac{-1}{\sqrt{r_{\mathrm{S}}}} \int^0_\delta r^{-1/2} \d r = 2\sqrt{\frac{\delta}{r_{\mathrm{S}}}}~,
\end{equation}
and thus the congruence of the causal curves covers the the solid angle of
\begin{equation}
  \Delta \Omega =\int ^{2\pi}_0\d \phi \int ^{\Delta \phi} _0 \d \theta ~\sin \theta = 2\pi(1-\cos \Delta \phi) \sim \pi \Delta \phi ^2 = \frac{4\pi \delta}{r_{\mathrm{S}}}~,
\end{equation}
which then means that the area covered by a single element of the anti-chain is
\begin{equation}
  S=\delta ^2\Delta \Omega = \frac{4\pi \delta^3}{r_{\mathrm{S}}} = 9\pi \tau ^2 ~.
\end{equation}

Therefore, for the Schwarzschild black hole, the lower bound of the density of the embedded anti-chain elements near the singularity as a function of proper time to the singularity is given by
\begin{equation}
  \rho _\mathrm{Schw} \equiv \frac{1}{\mathcal{L}_0 \cdot S}=\frac{2}{27\pi}\cdot \frac{1}{\tau ^3}~.
\end{equation}

\subsection{FLRW-flat}

Next we turn our attention to the spatially flat FLRW universe.
The metric in the comoving frame is given by
\begin{equation}
  g_{ab}=-\d \tau ^2 +a^2(\tau)\left( \d x^2+\d y^2+\d z^2 \right)~.
\end{equation}
We consider the dust-filled and radiation-filled cases:
\begin{gather}
  a_\mathrm{dust} (\tau) = \alpha \cdot \tau ^{2/3}~, \\
  a_\mathrm{rad} (\tau) = \beta \cdot \tau ^{1/2}~,
\end{gather}
where $\alpha$ and $\beta$ are some constants \cite{Wald}.

Null geodesics satisfy
\begin{equation}
  0=-1+a^2(\tau)\left( \frac{\d l}{\d \tau} \right)^2~,
\end{equation}
and thus the coordinate distance light travels in proper time $\tau$ becomes
\begin{equation}
  l=\int^\tau _0 \frac{\d \tau}{a(\tau)}=\begin{cases}
    \alpha^{-1}\int^\tau _0 \tau ^{-2/3} \d \tau = 3\alpha^{-1}\tau ^{1/3} & (\mbox{dust}); \\[5pt]
    \beta^{-1}\int^\tau _0 \tau ^{-1/2} \d \tau = 2\beta^{-1}\tau ^{1/2} & (\mbox{radiation}).
  \end{cases}
\end{equation}
A simple conversion from the coordinate distance to the proper distance along a straight line gives
\begin{equation}
  \mathcal{L}=a(\tau)\cdot l = \begin{cases}
    3\tau & (\mbox{dust}); \\
    2\tau & (\mbox{radiation}).
  \end{cases}
\end{equation}

Therefore, for the spatially flat FLRW universe, the lower bound of the density of the embedded anti-chain elements near the singularity (see the diagram on the right in Figure \ref{fig:signs} for its configuration) as a function of proper time to the singularity is given by
\begin{equation}
  \rho \equiv \frac{1}{\frac{4}{3}\pi \mathcal{L}^3} = \begin{cases}
    \frac{1}{36\pi} \cdot \frac{1}{\tau ^3} & (\mbox{dust}); \\[5pt]
    \frac{3}{32\pi} \cdot \frac{1}{\tau ^3} & (\mbox{radiation}).
  \end{cases}
\end{equation}

\subsection{Summary}

In the above calculations, we found
\begin{gather}
  \rho _\mathrm{Schw} = \frac{2}{27\pi}\cdot \frac{1}{\tau ^3}~, \\
  \rho_\mathrm{dust} = \frac{1}{36\pi} \cdot \frac{1}{\tau ^3}~, \\
  \rho_\mathrm{radiation} = \frac{3}{32\pi} \cdot \frac{1}{\tau ^3}~,
\end{gather}
all of which diverge when the embedded anti-chain is sent to the singularity as we expected.

Although these calculations do not serve as conclusive evidence, they suggest a correspondence between singularities and anti-chains consisting of either a countably infinite number of elements or an uncountable number of elements.
We, however, have strong evidence that countable anti-chains correspond to spatial hypersurfaces \cite{MRS}, and adopting the correspondence of singular anti-chains to singularities seems to be a reasonable choice.

\section{Causal set black hole}

Now that we have found how to describe singularities in the theory, it is a simple task to define causal set black holes.

\begin{figure}[h] \centering
  \begin{picture}(300, 100)
    \linethickness{1.5pt}
    \put(125,95){\line(1,0){50}}
    \put(125,10){\line(1,0){50}}
    \linethickness{0.4pt}

    \put(150,20){\circle*{3}}
    \put(130,35){\circle*{3}}
    \put(175,40){\circle*{3}}
    \put(125,50){\circle*{3}}
    \put(172,70){\circle*{3}} 
    \put(133,80){\circle*{3}}
    \put(152,75){\circle*{3}} 
    \put(165,90){\circle*{3}} 

    \qbezier(150,20)(150,20)(130,35)
    \qbezier(150,20)(150,20)(175,40)
    \qbezier(130,35)(130,35)(125,50)
    \qbezier(130,35)(130,35)(172,70)
    \qbezier(175,40)(175,40)(172,70)
    \qbezier(125,50)(125,50)(133,80)
    \qbezier(125,50)(125,50)(152,75)
    \qbezier(152,75)(152,75)(165,90)
    \qbezier(172,70)(172,70)(165,90)
    \qbezier(175,40)(175,40)(152,75)
    \qbezier(175,40)(175,40)(133,80)

    \qbezier(133,80)(115,90)(110,100)
    \qbezier(130,35)(115,10)(110,5)

    \qbezier(150,20)(150,20)(150,10)

    \qbezier(165,90)(165,90)(165,95)
    \qbezier(133,80)(133,80)(133,95)

    \qbezier(117,98)(160,55)(200,55)
    \put(204,53){$H_S$}

    \put(190,80){$B_S$}

  \end{picture}
  \caption{A causal set black hole and its horizon. The causal lines intersected by the curve across the diagram are the elements of the causal set black hole horizon, and the causal set black hole consists of the elements above the curve.}
  \label{fig:csbh}
\end{figure}
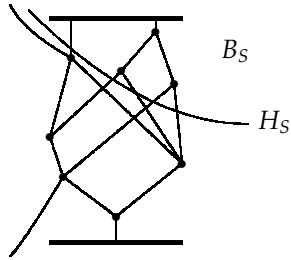

For a singular anti-chain, $S$, we define a causal set black hole, $B_S$, by
\begin{equation}
  B_S:=\{ x\in C~|~\forall ~p^+(x),~p^+(x)\cap S\neq \emptyset \}~,
\end{equation}
where $p^+(x)$ is a path starting at $x$ that is inextendible to its future. The condition means that $x\in C$ is inside the black hole if there is no future-directed maximal path from $x$ that can reach its future end without hitting the singular anti-chain. In other words, any such paths from $x$ inevitably reach the singular anti-chain with finite steps.
We can also define the causal set black hole horizon, $H_S$, as a set of links connecting the inside and the outside of the black hole:
\begin{equation}
  H_S:=\{(x, y)\in (C\setminus B_S)\times B_S~|~x\preceq y,~\mbox{and}~ |A(x,y)|=2 \}~.
\end{equation}
Figure \ref{fig:csbh} shows an example of a causal set black hole and its horizon.

\section{Discussion}

I briefly discuss some questions that singular anti-chains and causal set black holes may give rise to.

\subsection{Divergences due to the singularity}

First, we are interested in whether singular anti-chains introduce any diverging physical quantities.
The cardinality, which is interpreted as the volume of the corresponding spacetime, of a singular causet certainly diverges due to the uncountability of the singular anti-chain, but it would not have any observable effect since any local regions defined by Alexandroff intervals are still guaranteed to have finite cardinalities thanks to the local finiteness.

Another interesting observation concerns the Benincasa-Dowker action for a singular causet.
One can show that the two-dimensional BD action for a singular causet unavoidably diverges either positively or negatively depending on the structure of the causet near the singular anti-chain, but the four-dimensional BD action can converge\footnote{The detailed calculations will appear elsewhere.}. 
This might then put a restriction on the configuration of causal set black holes, and might help to construct physically realistic causal set black holes.

There is also an ambitious work in progress where the cosmological constant is used as the counter term to cancel the divergence in the BD action, thereby providing another new  way to calculate the cosmological constant in causal set theory.

\subsection{Structures of singularities and black holes}

\begin{figure}[h] \centering
  \begin{picture}(300, 140)
    \multiput(125,70)(0,50){2}{\multiput(0,0)(1,0){50}{\circle*{2}}}
    \multiput(125,70)(2,0){25}{\multiput(0,0)(0,4){13}{\line(0,1){2}}}

    \put(150,50){\circle*{3}}
    \put(100,100){\circle*{3}}
    \put(180,55){\circle*{3}}
    \put(200,90){\circle*{3}}
    \put(145,20){\circle*{3}}
    \put(50,105){\circle*{3}}
    \put(80,20){\circle*{3}}
    \put(220,70){\circle*{3}}
    \put(207,30){\circle*{3}}
    \put(270,114){\circle*{3}}

    \qbezier(150,50)(150,50)(150,70)
    \qbezier(180,55)(180,55)(160,70)
    \qbezier(180,55)(180,55)(200,90)
    \qbezier(220,70)(220,70)(200,90)
    \qbezier(200,90)(200,90)(165,120)
    \qbezier(150,50)(150,50)(145,20)
    \qbezier(145,20)(145,20)(100,100)
    \qbezier(145,20)(145,20)(50,105)
    \qbezier(100,100)(100,100)(130,120)
    \qbezier(207,30)(207,30)(220,70)
    \qbezier(207,30)(207,30)(180,55)
    \qbezier(80,20)(80,20)(50,105)
    \qbezier(80,20)(80,20)(100,100)
    \qbezier(270,114)(270,114)(220,70)
    \qbezier(270,114)(270,114)(275,140)
    \qbezier(50,105)(50,105)(60,130)
    \qbezier(220,70)(220,70)(260,40)
    \qbezier(145,20)(145,20)(150,10)
    \qbezier(80,20)(80,20)(70,10)
    \qbezier(207,30)(207,30)(220,10)

    \qbezier(117,80)(150,5)(184,80) \put(182,83){\small $H_\mathrm{in}$}
    \qbezier(70,120)(150,-50)(230,120) \put(228,123){\small $H_\mathrm{out}$}
  \end{picture}
  \caption{Two singular anti-chains connected by one-to-one causalities and two distinct inner and outer causal set black hole horizons.}
  \label{fig:timelike}
\end{figure}
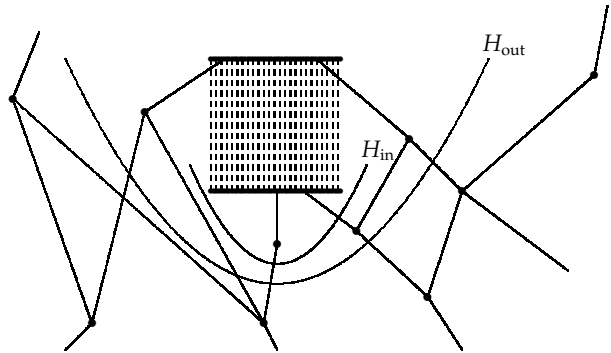

Another question is how we can describe different kinds of singularities with singular anti-chains.
Non-trivial spatial topology of a singularity, like a ring singularity, may simply be transferred to the corresponding singular anti-chain, but we have less idea about how to describe/distinguish time-like and space-like singularities.
One idea is to use a stack of singular anti-chains, connected by one-to-one causalities, to describe a time-like singularity.
In fact, two stacked singular anti-chains give us two distinct inner and outer causal set black hole horizons, as shown in Figure \ref{fig:timelike}, like the Reissner-Nordstr\"{o}m black hole.
Noticing that we can write down a metric with any number of horizons that is not necessarily a solution of the Einstein equations, we might expect the number of anti-chains stacked to be determined by the number of horizons.
However, we would need to establish the correspondence between a causet and a space-time beyond non-singular causets to draw any conclusion.

It is even more obscure to determine any characteristics of black holes, such as mass and charge, from causal set black holes.
Again, we do not have clear criteria for showing the correspondence between a spacetime and a singular causet since the sprinkling process does not work in the presence of a singular anti-chain.
This absence also makes it difficult to investigate causal set black holes in depth.

Furthermore, there is a fundamental question on the causal set black hole formation.
Currently, the best known dynamics of causal sets is the classical sequential growth model \cite{RS,Rideout}, with which causets with finite cardinality can only be produced, and there is no chance a singular anti-chain is created.

All the problems discussed here might become accessible when we obtain better understanding of the dynamics of causal sets and of the correspondence between a causet and a space-time, and we might have to wait until that time.

\subsection{Conclusion}

While singular anti-chains give us a lot of mysteries that cannot be resolved immediately, they may provide us a way to address interesting problems related to black holes in the future, such as Hawking radiation and black hole entropy, from the first principle of causal set theory assuming their interpretation proposed in this paper is correct. Although it will ultimately be determined with the development of the theory whether the interpretation is true or not, one thing that is certain is that singular anti-chains do exist in the theory, and we cannot ignore their existence in general.

\section*{Acknowledgments}
I am grateful to  Steven Carlip for his valuable suggestions during the course of the project. The calculations presented in Section \ref{sec:grcalc} was especially suggested by him.  I would also like to thank Joe Mitchell for his helpful comments.


\begin{thebibliography}{999}
\bibitem{BLMS}
  L. Bombelli, J. Lee, D. Meyer, and R. D. Sorkin, Space-time as a causal set, \emph{Phys. Rev. Lett.} \textbf{59} (1987) 521-524.
\bibitem{Sorkin}
  R. D. Sorkin, Causal sets: discrete gravity, arXiv:gr-qc/0309009 
\bibitem{ITR}
  R. Ilie, G. B. Thompson, and D. D. Reid, A numerical study of the correspondence between paths in a causal set and geodesics in the continuum, \emph{Class. and Quantum Grav.} \textbf{23} (2006) 3275-3286.
\bibitem{RW}
  D. Rideout and P. Wallden, Spacelike distance from discrete causal order, \emph{Class. and Quantum Grav.} \textbf{26} (2009) 155013.
\bibitem{BHS}
  L. Bombelli, J. Henson, and R. D. Sorkin, Discreteness without symmetry breaking: a theorem, \emph{Mod. Phys. Lett.} \textbf{A24} (2009) 2579-2587.
\bibitem{Johnston}
  S. Johnston, Feynman propagator for a free scalar field on a causal set, \emph{Phys. Rev. Lett.} \textbf{103} (2009) 180401.
\bibitem{Sverdlov}
  R. Sverdlov, Quantum field theory and gravity in causal sets, Ph.D. thesis (2009), arXiv:0905.2263.
\bibitem{BD}
  D. M. T. Benincasa and F. Dowker, The scalar curvature of a causal set, \emph{Phys. Rev. Lett.} \textbf{104} (2010) 181301.
\bibitem{DS}
  D. Dou and R. D. Sorkin, Black hole entropy as causal links, \emph{Found. Phys.} \textbf{33} (2003) 279-296.
\bibitem{RZ}
  D. Rideout and S. Zohren, Evidence for an entropy bound from fundamentally discrete gravity, \emph{Class. and Quantum Grav.} \textbf{23} (2006) 6195-6213.
\bibitem{HR}
  S. He and D. Rideout, A causal set black hole, \emph{Class. and Quantum Grav.} \textbf{26} (2009) 125015.
\bibitem{Wald}
  R. M. Wald, \emph{General relativity} (The University of Chicago Press, 1984) 
\bibitem{MRS}
  S. Major, D. Rideout, and S. Surya, Spatial hypersurfaces in causal set cosmology, \emph{Class. and Quantum Grav.} \textbf{23} (2006) 4743-4752.
\bibitem{RS}
  D. P. Rideout and R. D. Sorkin, Classical sequential grouwth dynamics for causal sets, \emph{Phys. Rev.} \textbf{D61} (1990) 024002.
\bibitem{Rideout}
  D. P. Rideout, Dynamics of causal sets, Ph.D. thesis (2001), arXiv:gr-qc/0212064. 
\end{thebibliography}
\end{document}